\documentclass[a4paper]{article}

\usepackage{INTERSPEECH2019}
\usepackage{amsmath}
\usepackage{cite}
\usepackage{subfigure}
\usepackage{amsmath,amssymb,amsfonts}
\usepackage{algorithm}
\usepackage{graphicx}
\usepackage{textcomp}
\usepackage{xcolor}
\usepackage{multirow}
\usepackage[export]{adjustbox}
\usepackage{breakcites}
\def\BibTeX{{\rm B\kern-.05em{\sc i\kern-.025em b}\kern-.08em
    T\kern-.1667em\lower.7ex\hbox{E}\kern-.125emX}}

\title{Exploration of Audio Quality Assessment and Anomaly Localisation Using Attention Models}
\name{Qiang Huang ~~~and~~~ Thomas Hain}
\address{
  Depart of Computer Science, University of Sheffield\\
  Sheffield, UK}
\email{{\{q.huang, t.hain\}}@sheffield.ac.uk}

\begin{document}


\maketitle

\begin{abstract}

Many applications of speech technology require more and more
audio data. Automatic assessment of the quality of the collected
recordings is important to ensure they meet the requirements of the related applications.
However, effective and high performing
assessment remains a challenging task without a clean reference.
In this paper, a novel model for audio quality assessment is proposed
by jointly using bidirectional long short-term memory and an attention mechanism.
The former is to mimic a human auditory perception ability to
learn information from a recording, and the latter is to further discriminate
interferences from desired signals by highlighting target related features.
To evaluate our proposed approach, the TIMIT dataset
is used and augmented by mixing with various natural sounds.
In our experiments, two tasks are explored. The first task
is to predict an utterance quality score, and the second is
to identify where an anomalous distortion takes place in a recording.
The obtained results show that the use of our proposed approach
outperforms a strong baseline method
and gains about 5\% improvements after being measured by 
three metrics, Linear Correlation Coefficient and Spearman’s Rank Correlation
Coefficient, and F1.

\end{abstract}

\section{Introduction}

Speech quality assessment aims to find and quantize the differences between original 
speech signals and the ones with variations.
There are two ways to assess speech quality: subjective and objective evaluation. 
The subjective evaluation  
is made by a listener's opinion
in terms of some pre-defined criterion, e.g., the mean opinion score (MOS).
The MOS is generally conducted by computing the arithmetic mean of 
all individual values on a predefined scale that a subject assigns to one's
opinion of the performance of a system quality 
\cite{streij1}.
As subjective evaluation may be time-consuming and expensive 
due to the need of human assessors,
objective quality evaluation has been widely used to predict the rating scores.
For objective evaluation, 
perceptual evaluation of speech quality (PESQ) \cite{pesq} 
can analyze an audio recording sample-by-sample after a temporal alignment of 
corresponding excerpts of reference and test signal. 
This means objective evaluation still requires a ``golden'' reference for each utterance to be evaluated, 
which considerably restricts 
the applicability of such assessment tools in real-world scenarios \cite{fu2018}.
Accordingly, it is highly desirable to develop a reliable assessment
model.

In recent years, due to the rapid development of deep neural networks,
some related technologies have been used for speech/voice quality assessment.
Spille et al., \cite{spille2017} used a deep neural network to predict
speech intelligibility. 
Soni et al.\cite{soni2016} applied a sub-band autoencoder to first learn features to be 
used by the following neural-network-based prediction model. 
Fu et al.\cite{fu2018} developed a non-intrusive speech quality evaluation model
to predict PESQ scores
using a BLSTM model on audio recordings.
Avila et al., \cite{shah2018} investigated the applicability
of three neural network-based approaches for non-intrusive audio quality assessment
based on mean opinion score(MOS) estimation.

In this paper, our task focuses on two aspects.
The first is to assess the quality of audio recordings
at utterance level, and the second is to
locate when an anomalous distortion takes place in the recording.
For this purpose, a frame-level based quality assessment architecture
using BLSTM \cite{schuster1997} is employed.
The structure is designed to compute the frame-level score, 
and then infer an utterance-level score. 
Moreover, by calculating frame-level scores and
detecting possible anomalous variations, 
anomaly regions can be thus located in a recording. 
To further increase the ability of quality assessment
against interferences,
an attention mechanism is employed.
The use of attention aims to allow focuses
to certain frames which are related to the target
set by users.
This means the target related information will probably be given a large weight, while
a small weight will probably be allocated to irrelevant features.
This is useful to discriminate interferences from desired signals,
and thus help to assess the quality of a recording and anomaly localisation.
The use of attention mechanisms has led to 
some state-of-the-art performances in different research fields, e.g.,
natural language processing \cite{bahdanau2014, denny2017, ashish2017},
speech recognition \cite{cortes2015,shinji2017, tian2019}, speaker
recognition\cite{bian2019, zhu2018, ind2019, qiang2019_1}, 
speech enhancement\cite{qiang2019_2, cas2016}. However, to our know knowledge,
more research in the use of attention on speech quality assessment is needed. 
The related details of the proposed architecture and how the
attention mechanism is used in our work will be presented in following sections.

The rest of paper is organised as follows: Section 2 depicts the details
of our proposed approach. In Section 3, the data used for model training
and evaluation and experiment setup are introduced. The related experimental
results and analysis are given in Section 4, and finally the conclusion is
drawn in Section 5.

\begin{figure}
\centering
    \includegraphics[width=72mm, height=72mm]{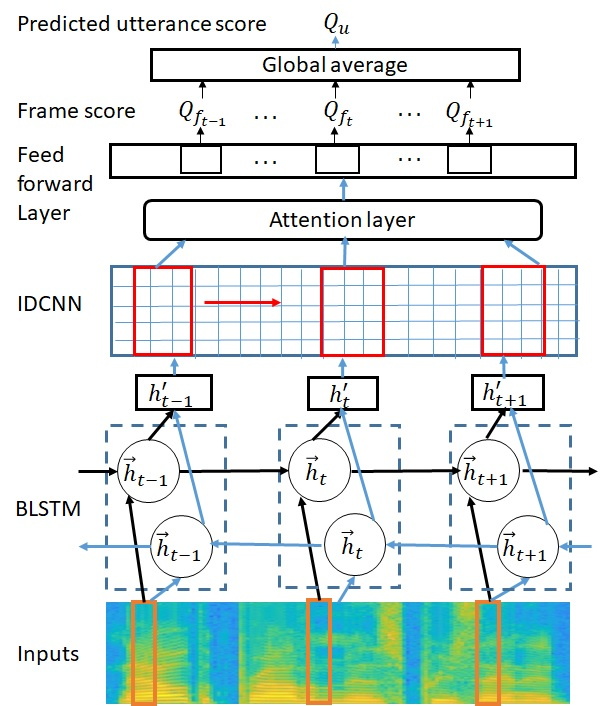}
\caption{Architecture of speech quality assessment and anomaly localisation using BLSTM, 1DCNN and attention model.}\label{fig:framework}
\end{figure}

\section{Proposed Architecture}\label{sec:framework}

Figure \ref{fig:framework} shows the proposed architecture using deep neural networks
and an attention model. Given the input spectrogram $\boldsymbol{S} \in \Re^{F\times T}$ 
of an utterance $\boldsymbol{U}$,
the proposed model aims to compute the quality $\boldsymbol{Q}_{\boldsymbol{f}_t} \in \Re^{1 \times T}$ of 
each frame $\boldsymbol{f}_t \in \Re^{F\times 1}$ of 
utterance $\boldsymbol{U}$, and then infer an utterance-level score 
$\boldsymbol{Q}_{\boldsymbol{U}} \in \Re^{1}$.
As shown in Figure \ref{fig:framework}, the proposed structure consists of
three parts. In the first part, a one-dimensional CNN (1DCNN) layer \cite{chollet2015} 
is cascaded with
a BLSTM layer to learns features using the contextual information in the time and frequency domains.  
An attention layer is used in the second part. The third part computes
the frame-level value $\boldsymbol{Q}_{\boldsymbol{f}_t}$ using a fully connected layer 
and finally infers
the utterance-level score $\boldsymbol{Q}_{\boldsymbol{U}}$ by averaging 
over all frames of utterance $\boldsymbol{U}$.

\subsection{Information Acquisition}\label{subsec:aaa}

The spectrogram $\boldsymbol{S}=(\boldsymbol{f}_{1} \cdots \boldsymbol{f}_{T})$ of
an input utterance $\boldsymbol{U}$ 
is processed by a BLTM layer and then by a one-dimensional CNN (1DCNN) layer.
\begin{eqnarray}
\footnotesize
    \boldsymbol{h}^{blstm} = \mathrm{BLSTM}~ (\boldsymbol{f}_{1..T}) \\
    \boldsymbol{h}^{1dcnn} = \mathrm{1DCNN}~ (\boldsymbol{h}^{blstm})
\end{eqnarray}
The BLSTM is an improvement over LSTM in that it captures both the previous 
timesteps (past features) and the future time steps (future features) 
via forward and backward states, respectively. 
It can be implemented by:
\begin{eqnarray}\label{equ:blstm1}
\footnotesize
\left.
  \begin{aligned}
    &\boldsymbol{\overrightarrow{h}}_{t} = \mathrm{LSTM} (\boldsymbol{f}_{t}, \boldsymbol{\overrightarrow{h}}_{t-1})\\
    &\boldsymbol{\overleftarrow{h}}_{t} = \mathrm{LSTM} (\boldsymbol{f}_{t}, \boldsymbol{\overleftarrow{h}}_{t+1})\\
    &\boldsymbol{h}^{blstm}_t = \boldsymbol{\overrightarrow{h}}_{t} + \boldsymbol{\overleftarrow{h}}_{t}
  \end{aligned}\right.
\end{eqnarray}
where the output of BLSTM layer $\boldsymbol{h}^{blstm} \in \Re^{L\times T}$ is formed
by concatenating the forward hidden state vector $\boldsymbol{\overrightarrow{h}}$ 
and the backward hidden state vector $\boldsymbol{\overleftarrow{h}}$, and $L$ is the vector dimension.  

The 1DCNN as described in \cite{1dcnn2019} makes use of:
\begin{equation}
    \boldsymbol{h}^{cnn}_{n} = \mathrm{K}_{n} \odot \boldsymbol{h}^{blstm}
\end{equation}
where $\odot$ denotes the convolution between the $n$th kernel $\mathrm{K}_n \in \Re^{F\times 3}$ ($n \in [1..N]$)
and the output of BLSTM $\boldsymbol{h}^{blstm}$. $\boldsymbol{h}^{cnn} \in \Re^{N \times T}$ represents the
output of 1DCNN layer.
The use of 1DCNN in this proposed architecture instead of 2DCNN 
is mainly because the 1DCNN has two advantages relating to feature 
extraction and computation efficiency \cite{1dcnn_2}. 
These advantages make it relatively easy to train and offer the small 
computational complexity while achieving state-of-the-art performance levels \cite{serkan2019}.

The use of the BLSTM is to mimic the human auditory perception system,
as a decision made by a human generally needs to consider the possible effects 
caused by contextual information, especially our final aim is to compute an
utterance-level score. 
Although the use of context information
might be helpful to the quality estimation of the current frame,
it may bring some negative impacts caused by 
the future or past frames if they are are corrupted by noise.
To more accurately predict frame quality and locate an anomaly,
the use of an attention model might be an effective way.

\subsection{Attention Model}\label{subsec:att}
The hidden state $\boldsymbol{h}^{blstm}$ of the BLSTM, computed by equation \ref{equ:blstm1},
is used as the input of an attention layer.
An attention matrix $A$ is formed by computing
the similarity between the hidden state $\boldsymbol{h}_{t}$
and $\boldsymbol{h}_{t'}$ corresponding to frame $\boldsymbol{f}_{t}$
and $\boldsymbol{f}_{t'}$ at timesteps $t$ and $t'$, respectively.
The attention mechanism is implemented as follows:

\begin{eqnarray}
\footnotesize
\left.
\begin{aligned}
    &h_{t,t'} = \mathrm{tanh}(\boldsymbol{h}^T_{t}\boldsymbol{W}_t + \boldsymbol{h}^T_{t'}\boldsymbol{W}_{t'} +\boldsymbol{b}_{t})\\
    &e_{t,t'} = \sigma(\boldsymbol{W}_{a}\boldsymbol{h}_{t,t'} + \boldsymbol{b}_{a})\\
    &\textbf{a}_t = \mathrm{softmax}(\textbf{e}_t)\\
    &l_t = \sum _{t'} a_{t,t'}\cdot \boldsymbol{h}_{t'}
\end{aligned}
\right.
\end{eqnarray}
where $\sigma$ is the element-wise sigmoid function, $W_t$ and $W_{t'}$
are the weight matrices corresponding to the hidden states $\boldsymbol{h}_{t}$
and $\boldsymbol{h}_{t'}$; $W_{a}$ is the weight matrix corresponding to 
their non-linear combination; $b_{t}$ and $b_{a}$ are the bias vectors.
The attention-focused hidden state representation $l_t$ of a frame
at timestep $t$ is given by the summation of the product of $\boldsymbol{h}_{t'}$
of all other frames at timesteps $t'$ and their similarity $a_{t,t'}$
to the hidden state representation $\boldsymbol{h}_t$ of the current frame.

\subsection{Loss function}\label{subsec:loss}
The score loss $L$ is defined by the summation of
an utterance-level loss ($L_{\textbf{u}}$) and the loss
($L_{\textbf{f}}$) averaged over all frames \cite{fu2018}: 
\begin{eqnarray}
\footnotesize
\left.
\begin{aligned}
    &L = L_{\textbf{u}} + L_{\textbf{f}}\\
    &L_{\textbf{u}} = MSE(Q_{u},Q'_{u})\\
    &L_{\textbf{f}} = \dfrac{1}{T}\sum_t(Q_{u}-Q'_{f_{t}})^2  \\
\end{aligned}
\right.
\end{eqnarray} 
where $Q_{u}$ is the target score of utterance $U$ and $Q'_{u}$
is its predicted value. $Q'_{f_{t}}$ represents the predicted quality score
of the $t$th frame.

\section{Experiment Setup}\label{sec:exp}

\subsection{Data}\label{subsec:data}
In our experiments, the TIMIT dataset \cite{timit92} was used as a comparison 
with the methods developed in \cite{fu2018}.
About 700 utterances in its training set
were used to train the proposed model,
and 143 utterances randomly selected from its test set
were used for evaluation.

Noise corrupted recordings are generated by
mixing the clean recordings with various natural sounds at five signal-noise ratio (SNR)
levels (-10dB, -5dB, 5dB, 10dB, 20dB). 
The noise signals used were from the the general noise portion
of the MUSAN dataset \cite{musan15}, which contains six hours of various natural sounds,
ranging from fax machine, car idling, thunder, wind, footsteps,
paper rustling, rain, and birdsong, etc. 

In all experiments, spectrograms are used as input features.
All of the audio streams are segmented using a 32-ms sliding window with a 16-ms shift.
A 512-point FFT was then used to convert each segment into a 257-dimension vector.

\subsection{Pseudo score}\label{subset:pss}

\begin{table}[tbh]
\footnotesize
\center
    \begin{tabular}{ |c|c| } 
 \hline 
 SNR (dB) & Pseudo Score \\ \hline  
 -10 & 1 \\ \hline
 -5 & 2 \\ \hline 
 5 & 4 \\ \hline
 10 & 5 \\ \hline
 20 & 7 \\ \hline
 original clean&8\\
 \hline
\end{tabular}
\caption{Pseudo scores defined in terms of SNR.}\label{tab:noise}
\end{table}
\vspace{-5mm}
In Table \ref{tab:noise}, a set of scores are defined by linking to SNR
values. 
This is to mimic the definition of scores used for MOS, but does not 
require human assessors' to mark each recording as the SNR value of a recording
can be set when precisely mixing the original recording with noise signals.
In addition, using a set of scores as assessment target might be able to mitigate
the impact caused by the use of noise corrupted target values e.g., using the estimated PESQ values
as targets in \cite{fu2018}.

The pseudo scores ($\{1,2,4,5,7\}$) are allocated to the
noise-corrupted utterances, whose SNR range from -10dB to 20dB with a 5dB shift.
In this experiment, the score of original clean speech is set to ``8''.

\subsection{Structure Configuration}\label{subsec:config}
Table \ref{tab:LC_ATT} shows the configuration of the proposed approach,
consisting of seven layers. In the first three layers,
the dimension of input frame vector
is 257, the output size of BLSTM is 200, and 250
kernels (size$=$3) used in 1DCNN.
The $Frame\_score$ layer
computes the frame-based prediction scores using a time-distributed
Dense layer, and the $Utterance\_score$ layer outputs
the utterance-level prediction using a GlobalAverage layer \cite{chollet2015}.

\subsection{Implementation}
Relying on the designed structure, experiments were conducted
using two proposed approaches and one baselines.
The first proposed approach LC$\_$ATT uses the structure
as presented in table \ref{tab:LC_ATT}, and in the second proposed
approach, L$\_$ATT, the same structure as LC$\_$ATT is employed without
the 1DCNN layer. The method developed in \cite{fu2018} is used as a baseline, which
did not use 1DCNN and the attention mechanism in 
comparison with our proposed approaches.
In experiments, RMSprop \cite{hinton2012} was used as an optimiser
and the initial learning rate was set to 0.001 with 0.95 decay every epoch. 

As both utterance-level and frame-level qualities are estimated from different layers,
as shown in Figure \ref{fig:framework},
regression 
instead of classification was used in our implementations.
This is also to compare with the baseline method (Baseline1)\cite{fu2018},
which used the same way to compute the utterance-level score.
In addition, the use of regression also enables the proposed
model to evaluate a recording, whose SNR is not listed in table \ref{tab:noise}, such as 15dB.
To compute F1, a threshold is set (threshold=7.1) in terms of the results obtained on the training data.

\begin{table}[tbh]
\footnotesize
\center
    \begin{tabular}{ ccc} 
 \hline 
 Layer name & Output shape & \#Param. \\ \hline
 InputLayer & (None, None, 257) & 0 \\ \hline 
 BLSTM      & (None, None, 200) & 286,400  \\ \hline
 1DCNN      & (None, None, 250) & 150,250  \\ \hline
 ATT        & (None, None, 250) & 16,065   \\ \hline
 Dense      & (None, None, 50)  & 12,550   \\ \hline
 Frame$\_$score&  (None, None, 1)  & 51       \\ \hline
 Utterance$\_$score& (None,1)      & 0        \\ 
 \hline
\end{tabular}
\caption{Configuration of the proposed network structure.}\label{tab:LC_ATT}
\end{table}

\subsection{Evaluation Metrics}\label{subsec:metrics}
The three metrics used to assess performance are
Linear Correlation Coefficient (LCC)\cite{frank1991},
Spearman’s Rank Correlation Coefficient (SRCC) \cite{wayne1990},
and F1 \cite{van1979}. The first two metrics are used to measure
the strength of the linear relationship between two variables.
The use of F1 is to measure the accuracy of
distinguishing the clean utterances from noise-corrupted 
ones.

\begin{figure*}\label{fig:gn}
\centering     
\subfigure[Baselin1]{\label{fig:gn_f1}\includegraphics[width=46mm, height=36mm]{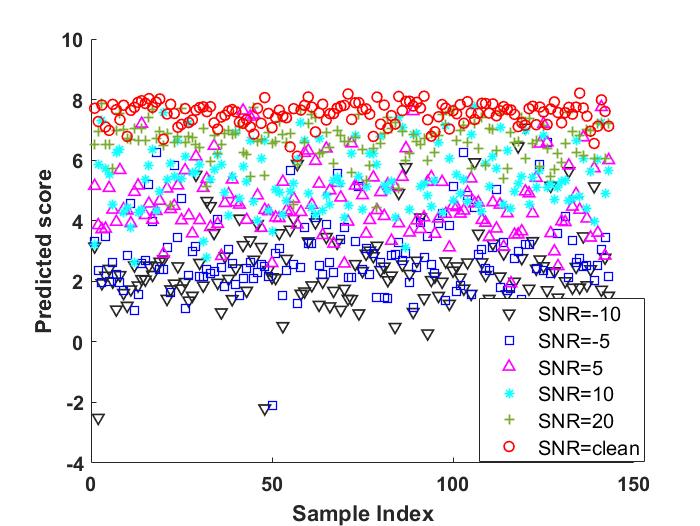}}
\subfigure[L$\_$ATT]{\label{fig:gn_f2}\includegraphics[width=46mm, height=36mm]{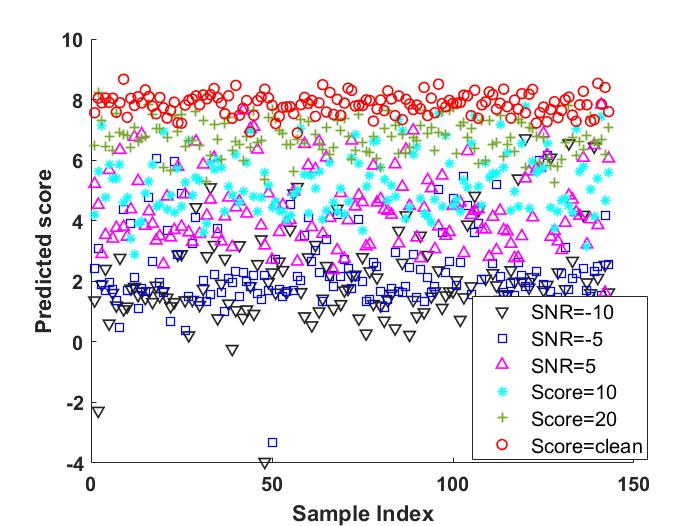}}
\subfigure[LC$\_$ATT]{\label{fig:gn_f3}\includegraphics[width=46mm, height=36mm]{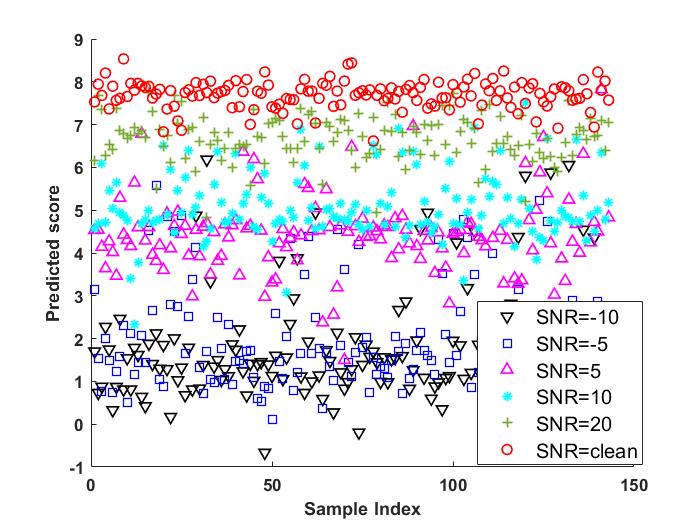}}
\caption{Predicted scores obtained on the test data using Baseline$\_$1, L$\_$ATT and LC$\_$ATT in noise conditions.}
\end{figure*}

\begin{figure*}
\centering     
\subfigure[Baseline1]{\label{fig:gner_f1}\includegraphics[width=42mm, height=36mm]{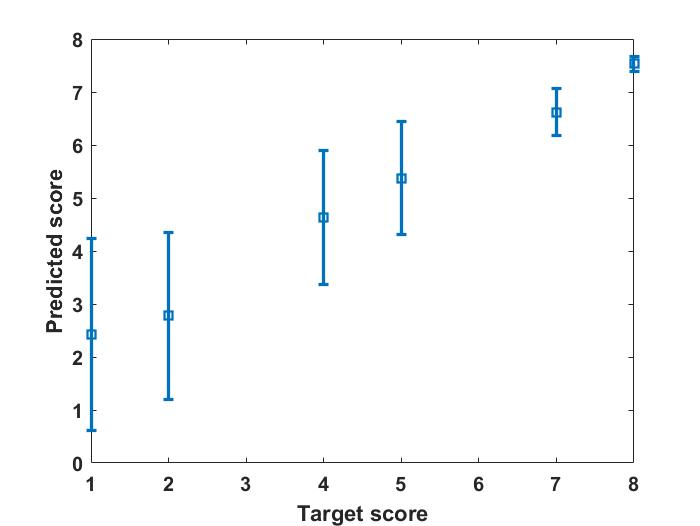}}
\subfigure[L$\_$ATT]{\label{fig:gner_f2}\includegraphics[width=42mm, height=36mm]{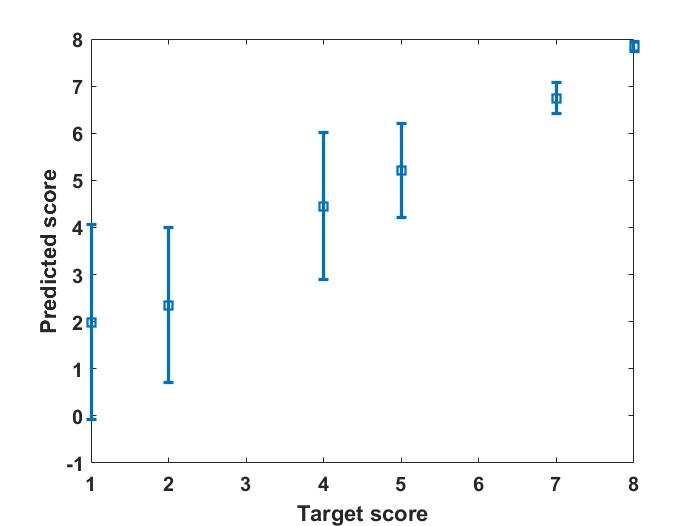}}
\subfigure[LC$\_$ATT]{\label{fig:gner_f3}\includegraphics[width=42mm, height=36mm]{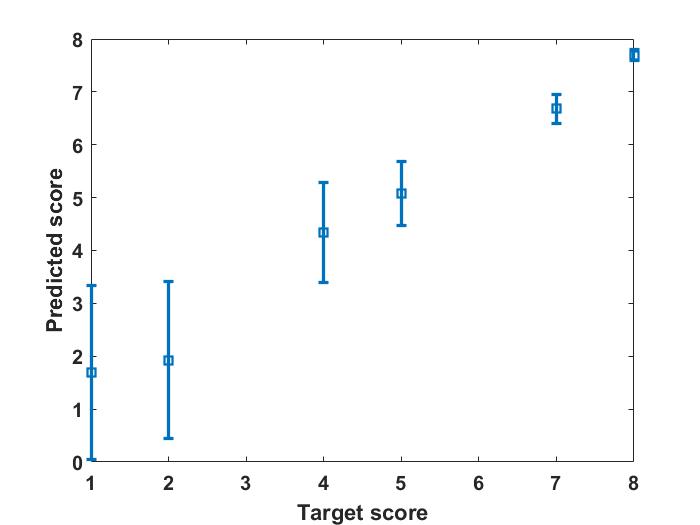}}
\caption{Mean and variance of predicted scores obtained on the test data using Baseline1, L$\_$ATT and LC$\_$ATT in noise conditions.}
\end{figure*}

\section{Results}\label{sec:results}
Figure 2 shows the predicted utterance-level quality
scores obtained using the baseline method (figure \ref{fig:gn_f1}) and our 
two approaches (figure \ref{fig:gn_f2} and \ref{fig:gn_f3})
in the condition of different distortions.
The x-axis in each figure denotes the test utterance index
and y-axis represents the predicted utterance-level scores. 
The three figures (figure \ref{fig:gn_f1}-\ref{fig:gn_f3})
show that
the predicted scores obtained using the proposed approaches
is closer to the target scores than Baseline1. 
Moreover, the corresponding statistics are also displayed
in \ref{fig:gner_f1}-\ref{fig:gner_f3}.
The error bars shown in the three figures
represent the mean values and the range of variance obtained
using the proposed approaches and the baseline method
in different conditions.
It can be found that the more signals are corrupted by noise, the higher
variances are generated.
This means it is hard to identify the quality of the audio signals in poor conditions.
In comparison with the baseline method, the use of our approach
can clearly reduce the predicted score variance
and the deviation between the target scores and the predicted scores.

Table \ref{tab:gn_metrics} lists LCC, SRCC, and F1 values obtained 
on the test data using
the baseline method and our proposed approaches. 
The results show that the use of our approaches can yield 
better performance than the baseline method.

\renewcommand{\arraystretch}{1.0}
\begin{table}[th]
\centering
\footnotesize
 \begin{tabular}{cccccc}
  \hline 
  \textbf{Method} & \textbf{LCC} & \textbf{SRCC} & \textbf{Precision} & \textbf{Recall} & \textbf{F1} \\  \hline
 
  Baseline1\cite{fu2018} & 0.876 &0.876 & 0.728 & 0.777 & 0.752 \\ \hline
  L$\_$1DCNN & 0.858 & 0.863& 0.926& 0.691& 0.792 \\ \hline
  L$\_$Att & 0.892 &0.894 & 0.957 & 0.709 & 0.815 \\ \hline
  LC$\_$Att & \textbf{0.919} &\textbf{0.914} & 0.927 & 0.781 & \textbf{0.848} \\ \hline
 \end{tabular}
 \caption{Metric values of LCC, SRCC, and F1 (larger is better) obtained on the test data corrupted by noise 
 using the baseline method and two proposed approaches.}\label{tab:gn_metrics}
 \end{table}

Since various natural sounds from MUSAN \cite{musan15} were 
used as noise to mix with clean signals, these noise signals
might take place at different time and have various effects in 
frequency domain.
To mitigate possible bias on 
LCC, SRCC, and F1 values, the experiments were repeated for eight times
and the average values are used as the final results.
To further compare the structure using an attention model and not using,
the third approach L$\_$1DCNN was also conducted.
It has the same structure as L$\_$ATT, but without the attention layer. 
The LCC and SRCC values listed in Table \ref{tab:gn_metrics} show that LC$\_$ATT 
can yield consistent
advantages over the baseline method, L$\_$1DCNN, and L$\_$ATT.

\begin{figure}[tbh]
\centering     
\includegraphics[width=79mm, height=63mm]{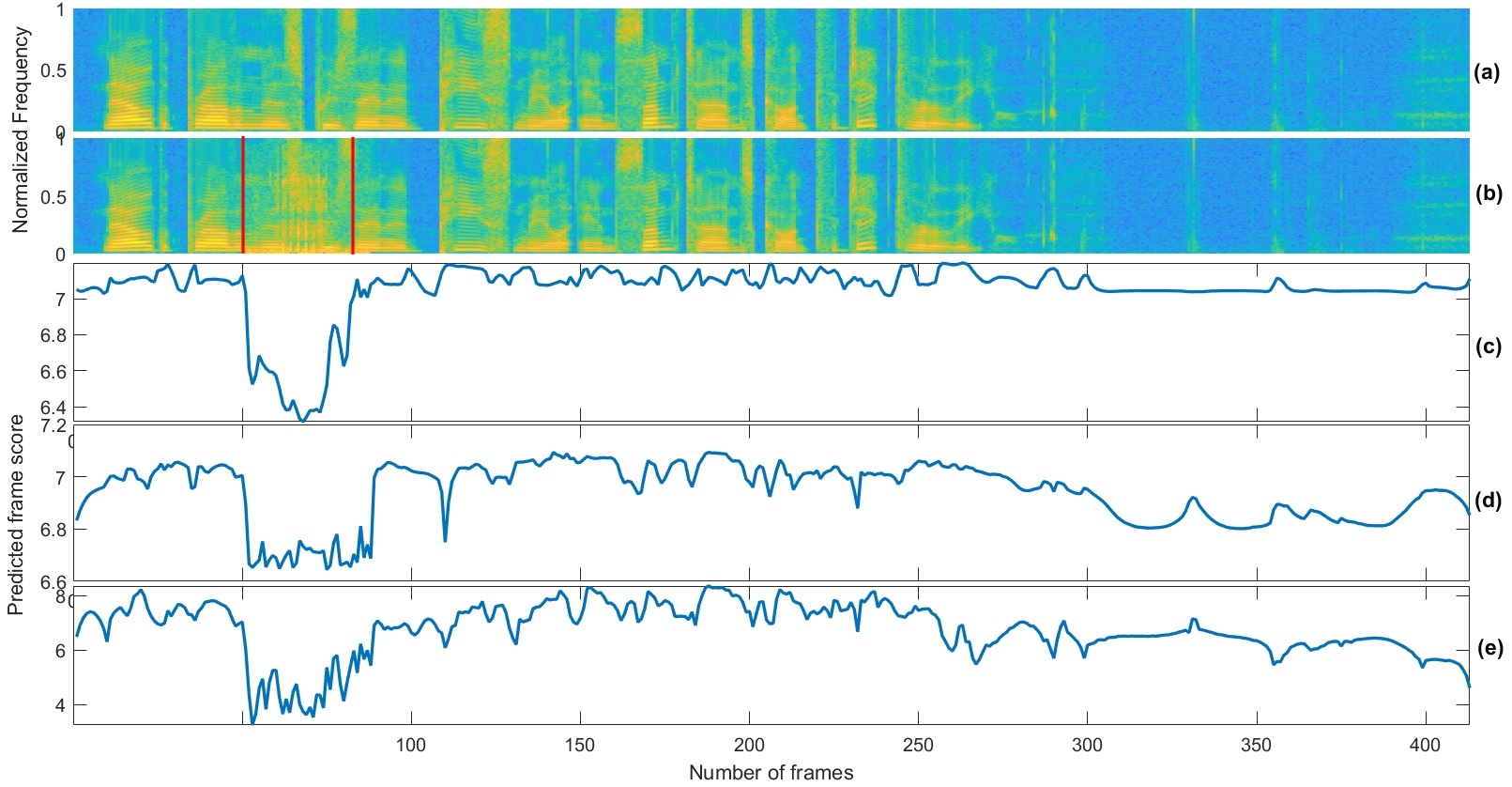}
\caption{Clean spectrograms (a), partially noise-corrupted spectrum(b), 
and the predicted frame-level scores obtained using LC$\_$ATT (c), L$\_$ATT (d), and Baseline1 (e).}\label{fig:spec_frame_score}
\end{figure}

In addition to predicting scores at utterance level,
identifying when an anomaly occurs in an audio recording
is also explored. 
We demonstrate how an anomalous distortion
can be located. 
In Figure \ref{fig:spec_frame_score}(a)-(e), the spectrogram of a recording
and its frame-level prediction scores are shown. 
From top to bottom, Figure \ref{fig:spec_frame_score}(a) is the spectrogram
of a clean utterance. The noise-corrupted spectrogram is shown in Figure \ref{fig:spec_frame_score}(b),
where an anomalous distortion (SNR=15dB) takes place from the 37$th$ frame to the 87$th$ frame, within
the range of two solid red lines.
The next three figures, Figure \ref{fig:spec_frame_score}(c)-(e), indicate
the frame-level prediction sores obtained using LC$\_$ATT, L$\_$ATT, and Baseline1, respectively.

It is clear that all of the three methods can find where the distortion is.
However, the use of Baseline1 generates a high score variation within the range where
the signal frames are corrupted by noise and outside. Compared to Baseline1, the use of LC$\_$ATT keeps a relative
smooth over all audio frames, by which the distortion range
is able to precisely located. This case is probably related to the use of 1DCNN and the attention mechanism.
The use of 1DCNN might mitigate the possible sudden variations by taking into account the
context information. The use of attention mechanism might be able to enlarge the difference
between clean signals and anomaly signals by highlighting the target related frame features.

\section{Conclusion and future work}\label{sec:conclusion}
A novel structure for audio quality assessment was
designed by using the BLSTM, one-dimensional convolutional neural networks and an attention mechanism.
It can assess the quality of audio recordings at utterance level
and identify the location of a distortion by 
computing frame-level scores. 
The obtained results, measured using three metrics, LCC, SRCC, and F1,
have shown that the use of attention model can
yield better performances than a strong baseline method
whether for utterance-level score prediction or for
anomaly distortion localisation.

In future, work in three aspects will be taken into account.
Firstly, some advanced neural network technologies, such as multi-head attention model, 
will be used to assess audio quality.
Secondly, the assessment technologies will be evaluated on large-sized
speech datasets and in various acoustic conditions.
Thirdly, the efficiency of assessment technologies will
be also evaluated to make it work in some practical applications.

\section{Acknowledgements}
This work was supported by Innovate UK Grant number 104264 MAUDIE.


\bibliographystyle{IEEEtran}

\bibliography{sqa}

\end{document}